\documentclass[runningheads]{llncs}
\usepackage[T1]{fontenc}
\usepackage{graphicx}
\usepackage{cite}
\usepackage{amsmath,amssymb,amsfonts}
\usepackage{algorithmic}
\usepackage{textcomp}
\usepackage{xcolor}
\usepackage{listings}
\usepackage{todonotes}
\usepackage{glossaries}
\usepackage{subcaption}
\usepackage{wrapfig} 
\usepackage[misc,geometry]{ifsym}

\usepackage[ruled]{algorithm2e}
\SetKwInOut{Input}{Input}
\SetKwInOut{Output}{Output\,}
\SetKwInOut{Data}{Data}
\SetKwProg{Tree}{Tree}{}{EndTree}

\setacronymstyle{long-short}

\newacronym{mcr}{MCR}{Maximum Cycle Ratio}
\newacronym{sps}{SPS}{Strictly Periodic Schedule}

\newacronym{cdfg}{CDFG}{Control Data Flow Graph}
\newacronym{cfg}{CFG}{Control Flow Graph}
\newacronym{dfg}{DFG}{Data Flow Graph}

\newacronym{df}{DF}{Data-Flow}
\newacronym{hsdf}{HSDF}{Homogeneous Synchronous Data-Flow}
\newacronym{sdf}{SDF}{Static Data-Flow}
\newacronym{csdf}{CSDF}{Cyclo-Static Data-Flow}
\newacronym{sdfap}{SDF-AP}{Static Data-Flow with Access Patterns}
\newacronym{sdfasap}{SDF-ASAP}{Static Data-Flow with Actors with Stretchable Access Patterns}
\newacronym{cp}{CP}{consumption pattern}
\newacronym{pp}{PP}{production pattern}

\newacronym{hof}{HOF}{Higher-order function}

\newacronym{rtl}{RTL}{Register-Transfer Level}
\newacronym{dag}{DAG}{Directed Acyclic Graph}

\newacronym{fsm}{FSM}{Finite-State Machine}
\newacronym{fifo}{FIFO}{FIFO}
\newacronym{fpga}{FPGA}{FPGA}

\newacronym{hls}{HLS}{High-Level Synthesis}
\newacronym{hdl}{HDL}{Hardware Description Language}
\newacronym{svd}{SVD}{Singular Value Decomposition}
\newacronym{dsp}{DSP}{Digital Signal Processor}
\newacronym{scop}{SCoP}{Static Control Parts}
\newacronym{pcp}{PCP}{Post Correspondence Problem}
\newacronym{qor}{QoR}{Quality of Results}
\newacronym{fu}{FU}{Functional Unit}

\newacronym{repl}{REPL}{Read-Evaluate-Print-Loop}
\newacronym{ast}{AST}{Abstract Syntax Tree}
\newacronym{ghc}{GHC}{Glasgow Haskell Compiler}

\newacronym{ssa}{SSA}{Static Single Assignment}

\newacronym{dct}{DCT}{Discrete Cosine Transform}
\newacronym{com}{CoM}{Center of Mass}

\newacronym{alm}{ALM}{Adaptive Logic Module}

\begin{document}

\title{High-Level Synthesis of Digital Circuits from Template Haskell and SDF-AP}

\author{H.H. Folmer \inst{1} \Letter \orcidID{0000-0003-3928-1658} \and
R. de Groote\inst{2} \and
M.J.G. Bekooij\inst{1}}

\authorrunning{Folmer et al.}

\institute{University of Twente, \\CAES: Computer Architectures for Embedded Systems, \\Enschede, Netherlands \\
\email{\{h.h.folmer, m.j.g.bekooij\}@utwente.nl}
\and
Saxion Hogeschool, \\Enschede, Netherlands \\
\email{e.degroote@saxion.nl}}

\maketitle

\begin{abstract}
Functional languages as input specifications for HLS-tools allow to specify data dependencies but do not contain a notion of time nor execution order.
In this paper, we propose a method to add this notion to the functional description using the dataflow model SDF-AP.
SDF-AP consists of patterns that express consumption and production that we can use to enforce resource usage.
We created an HLS-tool that can synthesize parallel hardware, both data and control path, based on the repetition, expressed in Higher-Order Functions, combined with specified SDF-AP patterns.

Our HLS-tool, based on Template Haskell, generates an Abstract Syntax Tree based on the given patterns and the functional description uses the Clash-compiler to generate VHDL/Verilog.

Case studies show consistent resource consumption and temporal behavior for our \gls{hls}.
A comparison with a commercially available HLS-tool shows that our HLS tool outperforms in terms of latency and sometimes in resource consumption.

The method and tool presented in this paper offer more transparency to the developer and allow to specify more accurately the synthesized hardware compared to what is possible with pragmas of the Vitis HLS-tool.

\end{abstract}

\section{Introduction}
A functional program describes a set of functions and their composition.
Referential transparency/side-effect free, a key feature of functional languages, not only makes formal reasoning easier but also prevents unwanted so-called false dependencies in the specification.
Because of this, functional specifications are inherently parallel, and as a consequence, there is no need for parallelism extraction\cite{edwards2006challenges}.
In the approach taken in this paper, we use a functional input language for hardware development.
Functional specifications can neither specify resource consumption nor temporal behavior, only data dependencies and (basic)operations.
To synthesize FPGA logic, the notion of both time and resource usage have to be introduced.
Due to a limited resource budget, one often has to perform a time-area trade-off.
Modelling time and resource consumption during the development stage could speed up the design process and indicate whether a certain design will meet the time and resource requirements.
Temporal modelling can also be used for latency and throughput analysis, and buffer size optimizations.

In this paper, we introduce a method to combine a functional description with consumption and production patterns according to the \gls{sdfap} model\cite{tripakis2011gluedesign}.
Given this specification, we generate a hardware design that is \textit{correct by construction}.
We have created an \gls{hls} tool that implements the methods proposed in this paper.
The tool uses the Clash-compiler and is part of the Haskell ecosystem~\cite{baaij2010clash}.
The Clash language is a subset of Haskell functional language and can be converted to VHDL or Verilog using the Clash-compiler.
The input specification is a purely functional description that only describes data dependencies.
This input description can be automatically converted to a \textit{fully combinational circuit} in VHDL or Verilog using the Clash-compiler.
The input description does not contain any clocked behavior and can be simulated and checked using the Haskell/Clash Interactive environment.
Often, due to resource constraints, the fully parallel description can not be synthesized as one combinational circuit because it does not fit on the FPGA or is too slow due to the length of the combinational path.
Our tool uses Template Haskell to convert the combinational Clash description into a clocked Clash description where hardware resources are shared over time, and registers and blockRAMs are introduced for (intermediate) storage.
This clocked description can also be simulated and checked in the Haskell/Clash Interactive environment.
This method allows for an \textit{iterative design} style because one can make a change to an individual node in the \gls{sdfap} graph and test the functional and temporal behavior in the same environment without entering the entire synthesis pipeline.

The proposed method uses access patterns from \gls{sdfap} to provide the engineer with a transparent way of performing the time-area trade-off.
The architecture generated is consistent with the given functional input description in combination with access patterns from the \gls{sdfap} graph.
The validity of these patterns can be checked by the compiler.
Invalid patterns will terminate the compilation process.
Changes to the access patterns consistently scale the resulting hardware architecture.
This consistency is required to provide an engineer with transparency on the consequences of his design choices.
Another advantage of the usage of the \gls{sdfap} model is that it allows for the usage of the analysis methods avaible\cite{ghosal2012sdfap}.

Many of the current state-of-the-art \gls{hls}-tools have C/C++ as input specification.
These languages are well known, have a large codebase, and a standard compiler already exists that can parse and check the code.
Imperative languages may allow direct control over storage, which was desirable for programming small embedded Von-Neumann devices but this could also apply to the synthesis of hardware architectures.
However, deriving true data dependencies and parallelism from sequential C++ code proves to be difficult\cite{8356004}.
To find only true data dependencies for a language that has pointers, one encounters the pointer aliasing problem which is undecidable\cite{landi1992undecidability,ramalingam1994undecidability}.
Therefore, the dependencies derived from the imperative input specification could contain false dependencies which limit the scheduling.
The current state-of-the-art \gls{hls}-tools like Vitis have introduced pragmas to allow the engineer to annotate the input specification to prevent false dependencies and influence the time-area trade-off\cite{VitisWeb}.
However, as opposed to the access patterns in our method, these pragmas can be ignored by the compiler, which makes the design process not transparent.
Our case studies show that sometimes the Vitis compiler ignores pragmas and generated inefficient designs.

In Section~\ref{sec:related_work} we discuss several temporal modelling and design techniques.
In Section~\ref{sec:sdfap} we explain the \gls{sdfap} model and in Section~\ref{sec:functional_sdfap_combination} we explain the basic idea of how we combined this model with a functional description.
The further workings of the \gls{hls} tool and design flow are described in Section~\ref{sec:toolflow}.
An example of code and some limitations of the tool are discussed in Section~\ref{sec:lloyds}.
Section~\ref{sec:case_study} contains three case studies, where we demonstrate the capabilities of our tool on a dot-product, Center of Mass (CoM) computations on images, and a 2D DCT.
The case studies demonstrate the effects of different access patterns and node decomposition on latency, throughput, and resource usage.
We also compare different versions in each case study with results from the Vitis \gls{hls} tool.
The input specification for Vitis is C++ code with pragmas to introduce parallelism.

\section{Related work}
\label{sec:related_work}
\subsection{High-Level Synthesis tools}
\gls{hls} is an active research topic and many major contributions have been made towards it.
Cong et al. give an overview of early \gls{hls}-tool developments\cite{5737854}.
They summarize the purpose and goals of \gls{hls} and indicate that there are many opportunities for further improvement.
Traditional \gls{hls}-tools use an imperative language as a behavioral input description and usually generate a \gls{cdfg} \cite{cong2006scheduling}.
The next step is to, given constraints, generate a structural \gls{rtl} description.
AutoPilot (a predecessor of the Vivado \gls{hls}-tool) is used to demonstrate the effectiveness of \gls{hls}, given a specification in C/C++.
They conclude that for C/C++ programs it remains difficult to capture parallelism which complicates both design and verification.

Sun et al. point out that resource sharing and scheduling are two major features in \gls{hls} techniques that the current \gls{hls}-tools still struggle with\cite{7428014}.
An important point that they make is that the \gls{hls}-tool obfuscates the relationship between the source code and the generated hardware, which in turn makes it hard to identify suboptimal parts of the code.
This non-transparency in the design flow is one of the aspects that this paper addresses.
Sun et al. also claim that for effective usage of \gls{hls}-tools, a rigorous understanding of the expected hardware is required.

Schafer et al. give a summary of multiple techniques used in the \gls{hls} process\cite{8847448}.
Controlling the process is typically done by setting different synthesis options, also called knobs.
The authors classify these knobs into three families:
The first knob is synthesis directives added to the source code in the form of comments or pragmas.
The second exploration knob is global synthesis options that apply to the entire behavioral description to be synthesized.
The last exploration knob allows users to control the number and type of \gls{fu}s.
Reducing the number of \gls{fu}s forces the \gls{hls} process to share resources, which might lead to smaller designs, albeit increasing the designs latency.

Lahti et al. present a survey of the scientific literature published since 2010 about the \gls{qor} and productivity differences between the \gls{hls} and RTL design flows\cite{8356004}.
The survey indicates that even the newest generation of \gls{hls}-tools do not provide as good performance and resource usage as manual RTL does.
Using an \gls{hls}-tool increases productivity by a factor of six, but iterative design is required.

Huang et al. present a survey paper of the scientific literature published since 2014 on performance improvements of \gls{hls}-tools\cite{Huang2020697}.
The survey contains a summary of both commercial and academic tools of which 13 of the 16 tools use C/C++ (subset) as input language.
In their conclusion, they state that the main work of optimization of \gls{hls}-tools is on improving the \gls{qor} but insufficient attention has been paid to improve the ease of use of the \gls{hls}-tools.

In our method, we use a functional language as input specification.
Whereas traditional \gls{hls}-tools need to generate structure from a behavioral description, our functional input specification already contains structure.
We are not required to obtain parallelism or derive only true dependencies because parallelism is implicit and the functional specification only contains true data dependencies\cite{edwards2006challenges}.
To find only true data dependencies for a language that has pointers, one encounters the pointer aliasing problem which is undecidable\cite{landi1992undecidability,ramalingam1994undecidability}.
Functional languages offer concepts like function composition, referential transparency, \gls{hof}s, and types, that provide a high level of abstraction\cite{baaij2010clash,Baaij}.

\subsection{Temporal models for hardware design}
There are several models available to analyze temporal behavior for hardware.
Lee and Messerschmitt introduced the \gls{sdf} model and it consists of nodes, edges and tokens, where edges have consumption and production values that specify the number of data elements (tokens) to be produced and consumed\cite{lee1987df}.
A node can fire when it has sufficient tokens, according to the consumption rates, on all of its input edges, and will produce tokens, according to the production rate, on all of its output edges at the end of its firing.

Horstmannshoff et al. glue high-level components together, usually from a library, by mapping \gls{sdf} to an \gls{rtl} communication architecture\cite{horstmannshoff1997dfhw, horstmannshoff1999dfhw}.
Multi-rate specifications are implemented as sequential components.
They present \gls{sdf} without a notion of time in the model and the presumption that every component, after an initialization, performs its calculation periodically.
Every component is slowed down to the period of the entire system using clock gating or adding registers (re-timing).
They use \gls{sdf} analysis techniques such as the repetition vector and topology matrix to determine how much every node (component) should be delayed.
A central control unit is introduced to provide the correct control signals to the datapath.

The GRAPE-II framework provides a sequential implementation of multi-rate dataflow graphs (\gls{sdf}, \gls{csdf}) with a distributed block control\cite{lauwereins1995grape,lauwereins1994grape}.
As input for hardware synthesis, it requires a VHDL description, then an engineer has to supply the tool with target-specific information, for example; clock frequency, resources.
Then it uses a set of different tools to do resource estimation, re-timing, assignment, routing, buffer minimization and scheduling.
Every tool bases its decisions on a comparison of performance estimates of various alternatives.
These estimates are obtained by calling the next tool in the script in estimate mode, in which a tool returns an estimate of the performance in an extremely short amount of time.

Ptolemy II is an open-source framework with an actor-oriented design that also supports VHDL code generation from dataflow specifications\cite{Eker2003ptolemy,leung2008vhdl,williamson1996synthesis, Williamson:M98/45}.
It implements a parallel structure for a multi-rate input graph.
From an \gls{sdf} graph, a \gls{dag} is constructed using a valid sequential schedule.
The \gls{dag} shows all the individual units of computation and the flow of data between them, and the hardware structure can be generated.
Sen and Bhattacharyya extend this technique providing an algorithm and framework to find the optimal application of data-parallel hardware implementations from \gls{sdf} graphs\cite{mainak2004parallelhardware}.

Chandrachoodan et al. provide a method for a hierarchical view of \gls{hsdf} graphs for \gls{dsp} applications\cite{chandrachoodan2004hierarchicalsdf}.
The new hierarchical node (block) has a delay of the worst case of all the paths from input to output inside.
The model uses \textit{timing pairs} that can be used to compute a \textit{constraint time}, which can be used to describe the "execution time" of a block.

\gls{sdf} is used in Jungs work to generate RTL codes for a hardware system including buffers and muxes\cite{jung2008optimized}.
They aim to generate a \textit{correct-by-construction} VHDL design from \gls{sdf} to accelerate both design and verification.
A node in their model corresponds to a coarse grain functional block such as an FIR filter and DCT.
It uses a centralized control structure and multi-rate specifications can be implemented parallel as well as sequentially or a hybrid.

Other approaches are based on the CAL language for HW/SW co-design\cite{janneck2008synthesizing, thavot2008dataflow}.
Actors are described in an XML format and transformed into a sequential program in \gls{ssa} from which hardware synthesis is performed.
Siret et al. use CAL in their \gls{hls} two-step approach in which they compile dataflow programs into hardware while keeping as many similarities as possible from the source and then letting the synthesis tool perform optimizations\cite{siret:hal-00763804}.

Kondratyev et al. propose an \gls{hls} scheduling approach in which they split the SystemC input specification into a \gls{cfg} and a \gls{dfg}\cite{kondratyev2012exploiting}.
The \gls{cfg} is constructed using conditionals, loops, and waits in the input specification.
The \gls{dfg} represents the operations with their data dependencies.
By using different scheduling methods they construct a mapping between both graphs and use a commercial \gls{hls} tool to synthesize hardware.

The methods mentioned above use dataflow to generate glue logic between components or analyze the throughput, latency, and the required buffer sizes of the application.
However, information from the model does not affect the implementation of the node itself.
In our approach, information from the model influences the hardware that is synthesized for the nodes, because we generate control and datapath from it, with the desired resource sharing and parallelism.

\section{SDF-AP}
\label{sec:sdfap}
\gls{sdf} and \gls{csdf} have the same underlying firing rule, which states that an actor can fire when there are enough tokens available on all its inputs.
Many hardware IP blocks, on the other hand, require that data arrive at specific clock cycles from the start of execution.
Suppose an actor requires $3$ tokens to be delivered in $3$ consecutive clock cycles, this cannot be expressed in \gls{sdf} or \gls{csdf} because the firing rule states that an actor waits until specified tokens arrive for this phase, not subsequent phases.
Therefore, designing hardware using these models can lead to inefficient or incorrect designs.
In \cite{tripakis2011gluedesign} this problem is further explained and an extension of \gls{sdf} is introduced called \gls{sdfap} is introduced.
This extension is elaborated in \cite{ghosal2012sdfap} and states that introducing actor stalling for example by disabling the clock to freeze the execution of a node is not a satisfactory solution due to the overhead.

\begin{figure}
  \centering
  \subfloat[DataFlow]{
    \centering
    \includegraphics[width=.4\textwidth]{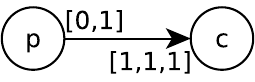}
    \label{fig:sdfap_ex_df}
    }
  \subfloat[Schedule]{
    \centering
    \includegraphics[width=.48\textwidth]{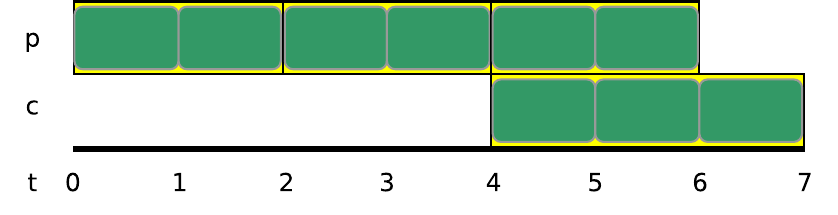}
    \label{fig:sdfap_ex_sc}
    }
  \caption{SDF-AP example}
  \label{fig:sdfap_ex}
  \vspace{-2em}
\end{figure}

\gls{sdfap} consists of a set of nodes and edges/channels with consumption and production patterns called \emph{access patterns}.
These patterns describe the number of tokens consumed or produced in each clock cycle of the node firing.
The execution time of a node denotes the number of clock cycles it takes to complete one firing.
\gls{sdfap} relies on strict pattern matching, which means that a node can only fire if it can be guaranteed that it will be able to complete all the phases.
The key difference between \gls{csdf} and \gls{sdfap} is that in \gls{csdf} it is allowed to have (stalling) time between phases of the firing, and in \gls{sdfap} this is not allowed.
An example of an \gls{sdfap} graph with a schedule is shown in Figure~\ref{fig:sdfap_ex}.
The actor \textit{p} produces, according to the production pattern $pp = [0,1]$, data after every second clock cycle of its firing.
The schedule (Figure~\ref{fig:sdfap_ex_sc}) shows $3$ consecutive firings of \textit{p}, starting at $t= 0, 2, 4$.
The actor \textit{c} requires that it can read $3$ tokens in $3$ consecutive clock cycles of its firing ($cp=[1,1,1]$).
At $t=4$ \textit{c} can fire because it is known upfront that it can complete its firing since the last token that is required will arrive at $t=6$.

\gls{sdfap} also has its shortcomings, which are explained in \cite{du2018sdflimitations, du2019sdfasap} and a solution is introduced, called \gls{sdfasap}, in which consumption patterns do not form a minimum requirement, but rather a maximum consumption pattern from which the real execution may be a stretched version.
Additional computation patterns are introduced to further specify the relation between tokens on the production and the consumption of a node.
This information can be used to stretch the patterns to allow for earlier execution of nodes and hence reduce the FIFO sizes for the edges.
For now, we focused on \gls{sdfap}, because even if earlier firings lead to smaller FIFO sizes, data has to be stored somewhere, and since we are generating both the entire control- and datapath those values will be stored inside a node.
Therefore it will increase the complexity of the controllers of both the FIFOs and nodes, it will increase the resource consumption of a node, but, it will decrease the FIFO sizes.
\gls{sdfasap} remains an interesting candidate for our solution because stretchable patterns can prevent the actor from stalling, especially for computations that require sliding windows.
Therefore, further implementation is future work.

\section{From functional description to SDF-AP}
\label{sec:functional_sdfap_combination}
In this section, we explore the key idea of combining \gls{sdfap} and the functional description for both design and analysability of hardware.
First, we show the general toolflow, after which we discuss the conformance relation between the model and hardware and what our tool will generate for each element in \gls{sdfap}.
After that, we discuss the basic idea of combining \gls{sdfap} patterns with a functional description to automatically generate a hardware architecture.
In section~\ref{sec:advantages} we describe the advantages of this approach.

\subsection{Template Haskell: The toolflow}
\label{sec:toolflow}
\begin{figure}
  \vspace{-1em}
  \centering
  \includegraphics[width=0.8\textwidth]{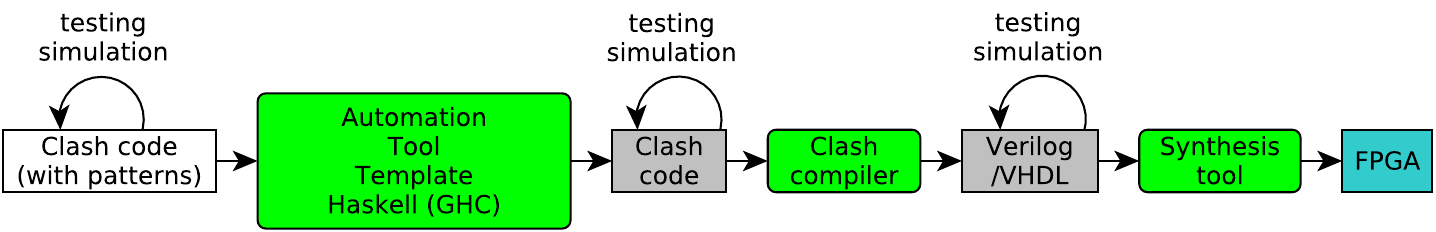}
  \caption{Toolflow}
  \label{fig:toolflow}
  \vspace{-2em}
\end{figure}

As mentioned in the introduction we use Template Haskell, an extension of the Haskell compiler, to transform the functional input description to a clocked clash description.
Template Haskell is the standard framework for doing type-safe, at compile-time metaprogramming in the \gls{ghc}.
It allows writing Haskell meta programs, which are evaluated at compile-time, and which produce Haskell programs as the results of their execution~\cite{sheard2002template}.
The Template Haskell extension allows us to analyze and change the \gls{ast} of a given description, through a process in which the \gls{ast} is extracted, modified, and afterward inserted back into the compilation process.
As an input for our tool, we have the functional specification combined with the specific access patterns for each input and output.
Invalid access patterns can be detected in the early stages of compilation and the engineer will be notified about the inconsistency in his specification.
Based on the patterns, our tool first generates a structure with partially predefined components.
The partially predefined components are for example a FIFO with a length that can be defined at compile-time or an input selector with a width that can be defined at compile-time.
Those components are now completed and linked together using the information from the patterns.
Then the tool generates an \gls{ast} from these now fully predefined components and inserts the \gls{ast} of the input description into it.
The new \gls{ast} now contains the control structure and the datapath from the functional description.
If there is repetition in the given \gls{ast}, using higher-order functions, then the tool can reuse the hardware of the repeating function according to the production and consumption patterns.
The amount of parallel hardware synthesized for the higher-order functions matches the patterns, this concept is further explained in Section~\ref{sec:basic_idea}.
The new "clocked" \gls{ast} is inserted back into the compilation process and VHDL or Verilog is generated.
An overview of this process is shown in Figure~\ref{fig:toolflow}.
Both the clash code and Verilog/VHDL are generated, they can be tested and simulated, but in practice, one would mainly test the input description, because the generated code is \textit{correct-by-construction}.
\subsection{Conformance relation}
\label{sec:conformance_relation}
This section provides the conformance relation between the \gls{sdfap} model and the hardware that is generated from it.
First, an example is presented to give an intuition of the conformance relation.
\subsubsection{example:}
\begin{figure}
  \centering
  \includegraphics[width=0.48\textwidth]{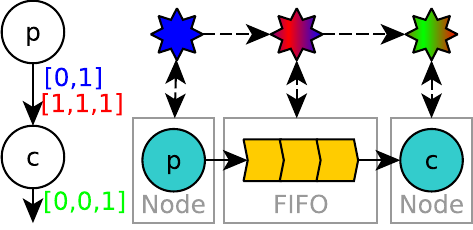}
  \caption{Conformance between model and hardware architecture}
  \label{fig:sdfap_ex_hw_vert}
  \vspace{-2em}
\end{figure}

An example is shown in Figure~\ref{fig:sdfap_ex_hw_vert} where the blue controller belongs to the \textit{p} node and controls when the node starts, when the result must be placed on the output, and signals the FIFO controller in what phase the node \textit{p} is in.
This node controller requires knowing the production pattern from the \gls{sdfap} model.
The FIFO controller controls when data must be placed in the FIFO and signals the controller of node \textit{c} (green/red) whether it can fire.
It needs the production (blue) pattern and the consumption (red) pattern from the model.
The node controller of \textit{c} (green/red) only knows the production (green) and consumption (red) pattern.

\subsubsection{edges:}
For every edge in the dataflow graph, a FIFO is generated.
Alongside the FIFO, a small controller is generated that checks whether, according to that specific edge, a node can fire.
This check is comparing the number of elements in the FIFO with the minimum number of elements required.
This minimum number required varies for each phase of the producing node.
Therefore, a list containing these minimums is calculated at compile-time using Algorithm~\ref{alg:fcs}.
A list for an edge between node A and B with $\mbox{PP} = [0,1,1]$ and $\mbox{CP} = [1,1,1,1,1]$ would be as follows:
If A is in its first phase of firing, then B can start if there are 2 or more elements in the FIFO because in the future A will produce 2 elements and B needs 4 elements in 4 clock cycles.
If A is in the second phase, then B can start if there are 2 or more elements in the FIFO because A will produce 1 element in this phase, and 1 in the next phase.
If A is in the third phase, then B can start if there are 3 or more elements in the FIFO because A will produce 1 element in this third phase.
So the list that keeps track of the minimum number of elements required in the FIFO before a node can fire according to that edge is [2,2,3] in this example.
Algorithm~\ref{alg:fcs} calculates such a list for a given production and consumption pattern.
For every firing (\textit{j}) the $\mbox{PPJ}$ is the remaining firing pattern.
For example, for the phase $j=1$ firing this $\mbox{PPJ} = [1,1]$.
0's are added to make the patterns equal, so $[1,1,0,0]$.
$\mbox{SPP}$ is a list that contains how many elements the node is going to produce in the future, so that is $[1,2,2,2]$ in our example for phase $j=1$.
$\mbox{SCP}$ contains the number of elements in total thus far required for every phase of the consumer, so for a $\mbox{CP}$ of $[1,1,1,1]$ that is $[1,2,3,4]$.
$\mbox{FC}_j$ is the maximum difference between $\mbox{SPP}$ and $\mbox{SCP}$, for the phase $j=1$, is $2$, hence $2$ elements are required in the FIFO if B can start its execution.
The above calculating is performed for every phase $j$ of the producing node.

In hardware, a FIFO can either be a blockRAM or a group of registers.
For now, the choice between blockRAM or registers depends on the size of the input.
The \gls{sdfap} model allows for varying integers as consumption and production, in our case we restrict the consumption and production patterns to only exist of $0$'s and $n$, where $n$ is the size of the incoming data.
All the patterns that belong to the same node have to have the same length.

\begin{algorithm}[]
  \caption{Algorithm to compute the minimum number of elements required in a FIFO before a node can fire.}
  \label{alg:fcs}
  \Input{\mbox{PP}, \mbox{CP}}
  \Output{\mbox{FC}, a list of minimum number of elements required in the FIFO for every phase of the producing node}
  \For {$j = 0$ to length($\mbox{PP}$)} {
    $\mbox{PPJ} = drop$ $j$ $\mbox{PP}$ \\
    add 0's to shortest pattern, so that $length(\mbox{CP}) == length(\mbox{PPJ})$\\
    \For {$i = 1$ to length($\mbox{CP}$)} {
      $ \mbox{SPP}_i = \sum_{0}^{i} \mbox{PPJ}_i $ \\
      $ \mbox{SCP}_i = \sum_{0}^{i} \mbox{CP}_i $
    }
    \vspace*{-2em}
    \begin{flalign*}
    &\mbox{FC}_j = \max_{i}{(\mbox{SPP}_i - \mbox{SCP}_i)} &
    \end{flalign*}
    \vspace*{-2em}
  }
\end{algorithm}
\subsubsection{nodes:}
A node represents a piece of hardware that can perform some task.
Data consumed and produced by the node are stored in the FIFOs that are generated from the edges.
However, a node can have its internal state stored inside.
Alongside the hardware of the node, there is a small controller generated that controls the operation.
It keeps track of the node phase counter and controls when the input and output are enabled, this is based on the consumption and production patterns.
The FIFO controllers signal whether, according to that specific edge, a node can fire.
The node controller receives these signals from the FIFO controllers and starts or continues the firing of the node.
It controls multiplexers for input and output values.

\subsection{The basic idea of combining \gls{sdfap} with a functional language}
\label{sec:basic_idea}
In a functional description, repetition is expressed using recursion or higher-order functions.
We currently focus on higher-order functions due to their expressiveness of structure.
The basic idea is that we combine consumption and production patterns from \gls{sdfap} with functions from the specification.
Combining the repetition, expressed using higher-order functions, with patterns from \gls{sdfap}, we can generate a hardware architecture with a time-area trade-off automatically.
This principle is explained in the following example using a dot-product specification in Haskell (Listing~\ref{lst:dotp}), with two higher-order functions (\textit{foldl1} and \textit{zipWith}).
\textit{foldl1} is called a foldable function because it combines all input values to a single output.
According to the type definition (line 1-3), the \textit{dotp} function receives $2$ vectors of $6$ values and produces a single value.

\begin{figure}
  \vspace{-2em}
  \centering
  \begin{subfigure}{.20\textwidth}
    \centering
    \includegraphics[width=\textwidth]{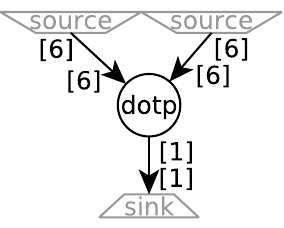}
    \caption{\scriptsize{DataFlow}}
    \label{fig:dotp_df}
    \vspace{-.5em}
  \end{subfigure}
  \begin{subfigure}{.15\textwidth}
    \centering
    \includegraphics[width=\textwidth]{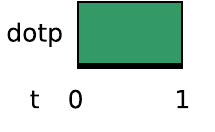}
    \caption{\scriptsize{Schedule}}
    \label{fig:dotp_sc}
    \vspace{-.5em}
  \end{subfigure}
  \begin{subfigure}{.60\textwidth}
    \centering
    \includegraphics[width=\textwidth]{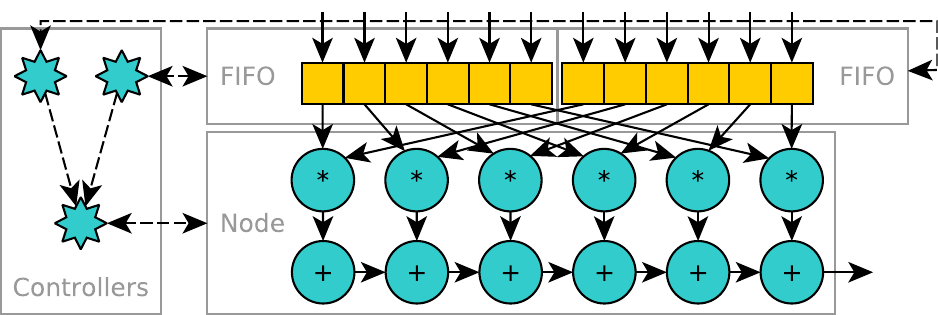}
    \caption{\scriptsize{Hardware}}
    \label{fig:dotp_hw}
    \vspace{-.5em}
  \end{subfigure}
  \caption{Function \textit{dotp} with $cps=[6]$ and $pp=[1]$}
  \label{fig:dotp}
  \vspace{-2em}
\end{figure}

\begin{lstlisting}[label={lst:dotp}, caption={\textit{dotp} function}]
dotp :: Vec 6 (Unsigned 8)
     -> Vec 6 (Unsigned 8)
     -> Unsigned 8
dotp xs ys = o
  where o  = foldl1 (+) 0 ws
        ws = zipWith (*) xs ys
\end{lstlisting}
In the first design iteration, we can have a single \gls{sdfap} actor representing the \textit{dotp} function with both consumption patterns $cp = [6]$ and the production pattern $pp = [1]$ (Figure~\ref{fig:dotp_df}).
The schedule in Figure~\ref{fig:dotp_sc} shows that the system takes $1$ clock cycle to complete its computation.
Our tool receives both the \textit{dotp} function (Listing~\ref{lst:dotp}) and the consumption and production patterns (as shown in Listing~\ref{lst:dotp_n}) and generates the hardware architecture as shown in Figure~\ref{fig:dotp_hw}.
The patterns are given in a list of tuples containing production and consumption pattern of every edge (pp,cp).
In Listing~\ref{lst:dotp_n} the tuple for both incoming edges is ([6],[6]), so the production pattern for both edges is the same as the consumption pattern.
The hardware consists of, as described in the conformance relation, two FIFOs for incoming data, $6$ multipliers, $6$ adders, and controllers for both FIFOs and a controller for the \textit{dotp} node.
The resource consumption is consistent with the access patterns.
\begin{lstlisting}[label={lst:dotp_n}, caption={\textit{dotp} in the Template Haskell function}]
$(tool 'dotp [([6],[6]),([6],[6])] [1])
\end{lstlisting}

\begin{figure}
  \vspace{-2em}
  \centering
  \begin{subfigure}{.20\textwidth}
    \centering
    \includegraphics[width=\textwidth]{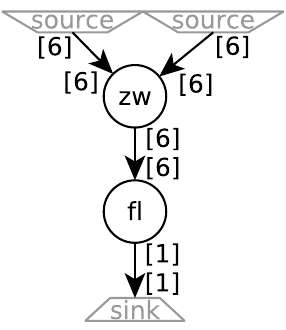}
    \caption{\scriptsize{DataFlow}}
    \label{fig:zwfl_6661_df}
    \vspace{-.5em}
  \end{subfigure}
  \begin{subfigure}{.15\textwidth}
    \centering
    \includegraphics[width=\textwidth]{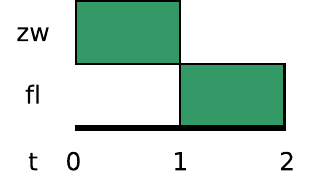}
    \caption{\scriptsize{Schedule}}
    \label{fig:zwfl_6661_sc}
    \vspace{-.5em}
  \end{subfigure}
  \begin{subfigure}{.60\textwidth}
    \centering
    \includegraphics[width=\textwidth]{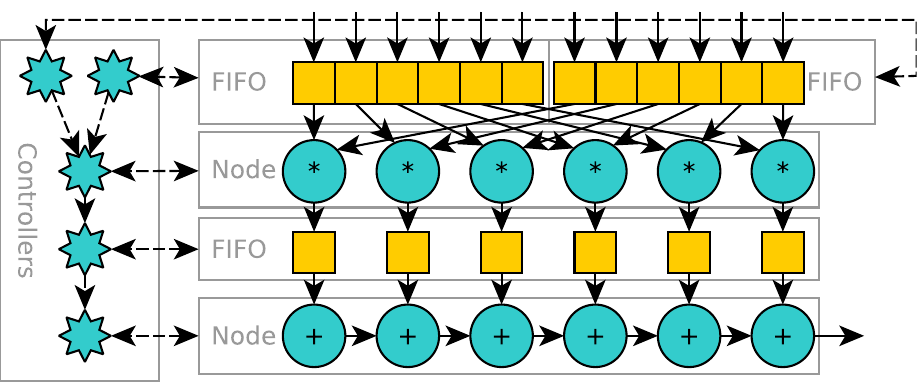}
    \caption{\scriptsize{Hardware}}
    \label{fig:zwfl_6661_hw}
    \vspace{-.5em}
  \end{subfigure}
  \caption{Function \textit{dotp} with separate nodes for \textit{zipWith} and \textit{foldl1}}
  \label{fig:zwfl_6661}
  \vspace{-2em}
\end{figure}

In the second design iteration, we can examine the \textit{dotp} function and model both higher-order functions (\textit{foldl1} and \textit{zipWith}) as separate \gls{sdfap} actors (Figure~\ref{fig:zwfl_6661_df}).
The consumption and production patterns of the \textit{zw} node are $cp_{zw} = [6]$, and $pp_{zw} = [6]$.
The consumption and production patterns of the \textit{fl} node are $cp_{fl} = [6]$, and $pp_{fl} = [1]$.
The hardware (Figure~\ref{fig:zwfl_6661_hw}) that is generated from these patterns and the function description of both higher-order functions consists of $6$ multipliers, $6$ adders, and FIFOs on the input edges of both nodes.
According to the schedule (Figure~\ref{fig:zwfl_6661_sc}), the entire system takes $2$ clock cycles before producing the result.
Introducing an additional edge results in an additional FIFO, this is fully transparent for the engineer and the resource consumption shows consistent behavior.

\begin{figure}
  \vspace{-2em}
  \centering
  \begin{subfigure}{.18\textwidth}
    \centering
    \includegraphics[width=\textwidth]{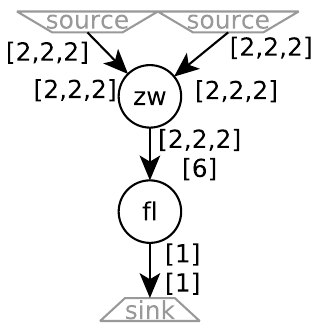}
    \caption{\scriptsize{DataFlow}}
    \label{fig:zwfl_2261_df}
    \vspace{-.5em}
  \end{subfigure}
  \begin{subfigure}{.20\textwidth}
    \centering
    \includegraphics[width=\textwidth]{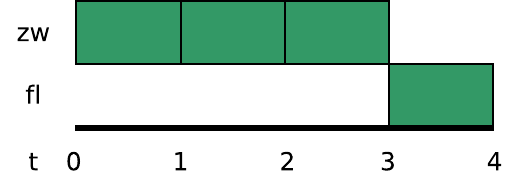}
    \caption{\scriptsize{Schedule}}
    \label{fig:zwfl_2261_sc}
    \vspace{-.5em}
  \end{subfigure}
  \begin{subfigure}{.60\textwidth}
    \centering
    \includegraphics[width=\textwidth]{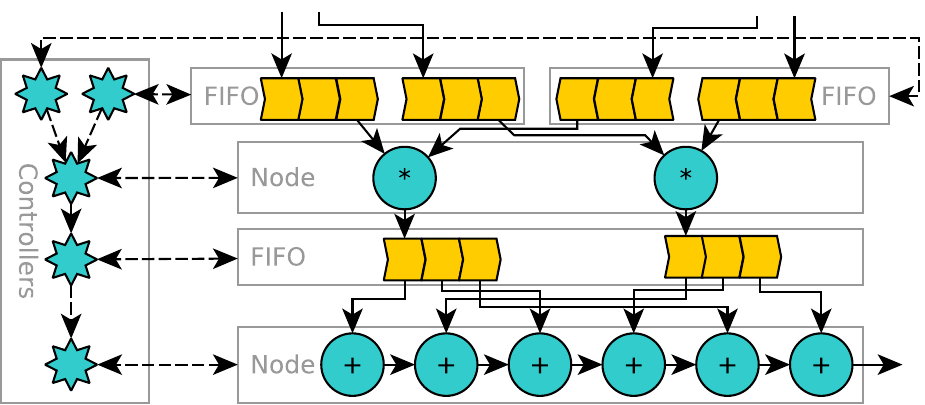}
    \caption{\scriptsize{Hardware}}
    \label{fig:zwfl_2261_hw}
    \vspace{-.5em}
  \end{subfigure}
  \caption{Function \textit{dotp} with modified patterns for \textit{zw}}
  \label{fig:zwfl_2261}
  \vspace{-2em}
\end{figure}

In the third design iteration, we can change the consumption and production patterns of the \textit{zw} actor to $cp_{zw} = pp_{zw} = [2,2,2]$ (Figure~\ref{fig:zwfl_2261_df}).
There are a couple of different permutations possible on the patterns that would result in a feasible architecture, those permutations are $[1,1,1,1,1,1], [2,2,2], [3,3]$.
These are the divisors of the original pattern.
For now, we discard all the remaining permutations because in hardware it introduces control-overhead if we allow different integers in the same pattern.
The resulting hardware (Figure~\ref{fig:zwfl_2261_hw}) consists of $2$ multipliers, $6$ adders, and FIFOs on the input edges of both nodes.
According to the schedule (Figure~\ref{fig:zwfl_2261_sc}), the entire system takes $4$ clock cycles.

\begin{figure}
  \vspace{-2em}
  \centering
  \begin{subfigure}{.20\textwidth}
    \centering
    \includegraphics[width=\textwidth]{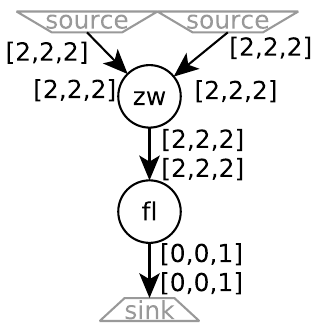}
    \caption{\scriptsize{DataFlow}}
    \label{fig:zwfl_2221_df}
    \vspace{-.5em}
  \end{subfigure}
  \begin{subfigure}{.25\textwidth}
    \centering
    \includegraphics[width=\textwidth]{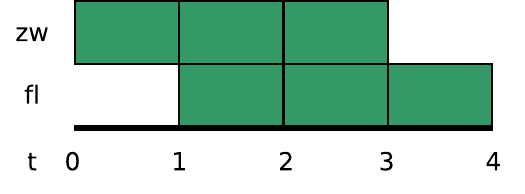}
    \caption{\scriptsize{Schedule}}
    \label{fig:zwfl_2221_sc}
    \vspace{-.5em}
  \end{subfigure}
  \begin{subfigure}{.50\textwidth}
    \centering
    \includegraphics[width=\textwidth]{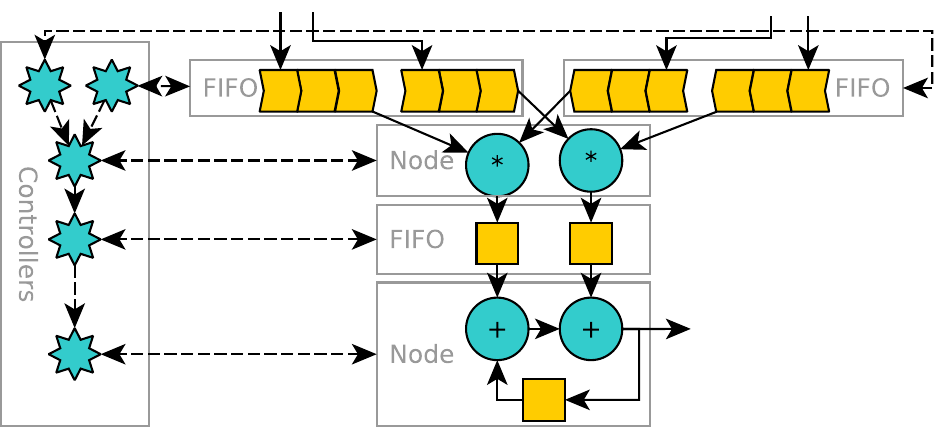}
    \caption{\scriptsize{Hardware}}
    \label{fig:zwfl_2221_hw}
    \vspace{-.5em}
  \end{subfigure}
  \caption{Function \textit{dotp} with modified patterns for both \textit{zw} and \textit{fl}}
  \label{fig:zwfl_2221}
  \vspace{-2em}
\end{figure}

In the previous scenario, it can be seen that it is very inefficient to leave the \textit{fl} function untouched.
The $6$ adders in the \textit{foldl1} part of the system are idle in $3$ of the $4$ clock cycles.
Therefore, bundling the production pattern of the \textit{zw} node to the \textit{fl} node results in a much more efficient architecture.
The consumption pattern of the \textit{fl} node then becomes $cp_{fl} = [2,2,2]$ and the production pattern $pp_{fl} = [0,0,1]$.
Changing these patterns results in an architecture that consists of $2$ multipliers, $2$ adders, and FIFOs on input edges (Figure~\ref{fig:zwfl_2221_df} and \ref{fig:zwfl_2221_hw}).
The schedule (Figure~\ref{fig:zwfl_2221_sc}) now shows that the entire system also takes $4$ clock cycles.
This is because the \textit{fl} node can start as soon as the \textit{zw} node has finished its first firing phase.
The control logic generated by the tool facilitates this schedule automatically.
We also introduced the first optimization; if an edge has the same production pattern as the consumption pattern we remove the FIFO controller and introduce a pipeline register.
Both resource consumption and the schedule of the architecture match the expectations of the graph with access patterns, making the design process transparent.
The resource consumption in \gls{dsp}s scales consistently with the change in patterns.

\subsection{The advantages}
\label{sec:advantages}
There are several advantages of method and tool to combine a functional description with \gls{sdfap}:
\begin{itemize}
  \item The generation of control and datapath is automated, and therefore lifts the burden of the engineer.
  \item The time-area trade-off is transparent.
  By tuning the consumption and production patterns including the functional description, the engineer steers both the timing and the structure of the generated architecture.
  \item The engineer can steer the time-area trade-off by describing the functionality using specific higher-order functions.
  Often, functionality can be expressed using different higher-order functions.
  If an engineer knows beforehand that these higher-order functions are the first place where a time-area trade-off can be made, he can choose to use specific higher-order functions to steer the direction of this trade-off.
  \item The typechecker of functional languages can be used to check and verify the input specification with access patterns.
  This typechecker is also part of the Haskell ecosystem.
  \item Iterative design is possible due to the analysis and simulation techniques of dataflow and functional languages.
  Haskell comes with an interactive \gls{repl} that allows simulation of functional behavior.
  An engineer can use several analysis techniques from the \gls{sdfap} model to determine throughput, latency, buffer sizes, and bottlenecks and change the input specification before entering the remaining design flow.
  \item The possibility to introduce hierarchy.
  Production and consumption patterns can be bundled, multiple nodes with bundled patterns can be modelled as one node, which will reduce the search space for automation in the future.
  Bundling not only allows for hierarchy but also excludes irrational design pattern combinations that lead to inefficient hardware architectures.
  Besides bundling it is also possible to check how local changes to the design influence the design as a whole.
\end{itemize}
\subsection{The current limitations}
There are several limitations of the proposed method and tool:
\begin{itemize}
  \item Due to the choice for distributed local controllers, there is hardware control overhead introduced at every edge and node of the \gls{sdfap} model.
  The overhead introduced by the node controllers is smaller than the FIFO controllers, since they only switch multiplexers on the input and output of the node based on signals received from FIFO controllers, and count phases.
  The FIFO controllers have to count the number of elements in the FIFO and if enough elements are presents, signals the node controller.
  Since production and consumption patterns are known at compile-time, many calculations can be performed at compile-time and therefore reducing the size of the circuitry.
  Still, if the nodes are very small components, for example, one adder, then the controller overhead is relatively large.
  \item Only the repetition expressed in higher-order functions allows for an automatic time-area trade-off based on patterns.
  \item There is no algorithm yet that sets production and consumption patterns, the selection of these patterns is still completely up to the engineer.
  Future work remains to search the design space automatically and find access patterns that result in an architecture that both satisfies area and time constraints.
  \item If the tool receives a folding function at the root of the \gls{ast}, it will automatically determine the state variables if the hardware needs to share its resources over time.
  However, if there is a folding function somewhere inside the \gls{ast}, and not at the root, the tool is unable to determine the internal state required and generate an architecture.
  For example, if the \textit{dotp} function from Listing~\ref{lst:dotp} is given to our tool, then it is unable to determine that there is a \textit{foldl1} function inside it.
  Hence our tool is not able to determine the internal state required for an architecture.
  This limitation is further demonstrated in the Lloyds case study (See Section~\ref{sec:lloyds}) and is planned to be resolved in the future.
  \item Resources between different higher-order functions will not be shared by this method alone.
  For example, the function \textit{foo} from Listing~\ref{lst:shortcomming_reuse} uses two times the higher-order function \textit{zipWith}, but the tool is currently unable to share the resources between those two higher-order functions.
  This means that the minimum number of multipliers required is $2$ (one for each zipWith).
\end{itemize}
\vspace{-1em}
\begin{lstlisting}[label={lst:shortcomming_reuse}, caption={Function with multiple \gls{hof}s}]
foo xs ys zs = (o1, o2) where
  o1  = zipWith (*) xs ys
  o2  = zipWith (*) xs zs
\end{lstlisting}

\section{Node decomposition}
\label{sec:lloyds}
\begin{wrapfigure}{r}{0.31\textwidth}
  \vspace*{-2em}
  \centering
  \includegraphics[width=0.30\textwidth]{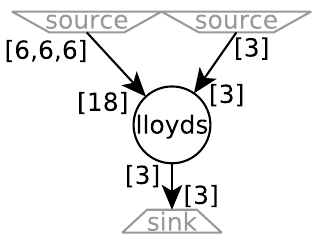}
  \caption{Lloyds 1 node}
  \label{fig:lloyds_1n_df}
  \vspace*{-2em}
\end{wrapfigure}

In the section, we demonstrate the effects of decomposing a single node into multiple nodes on resource usage and show the code necessary to specify the \gls{sdfap} graph in our \gls{hls} tool.
Decomposing a single node means introducing extra edges, and hence extra FIFOs in the architecture.
To demonstrate this we use Lloyds algorithm\cite{lloydsAlgorithm} that finds the center of each set of Euclidean spaces and re-partitions the input to the closest center.
It consists of iterating through these two steps: assigning data points to a cluster and then centering the cluster point.
For demonstration purposes, the input is $18$ different points that the hardware needs to cluster using Lloyds algorithm.
The synthesis results of all the different versions are shown in Table~\ref{table:lloyds_resources}.

\begin{table*}
  \vspace{-2em}
\centering
\caption{Resources usage for different versions of the Lloyds algorithm}
\begin{tabular}{l||c|c|c|c|c||c|c|c|c|}
Nodes             & 1 (Fig.\ref{fig:lloyds_1n_df})  & 3 (Fig.\ref{fig:lloyds_3n_df})  & 4 (Fig.\ref{fig:lloyds_4n_df})  & 5 (Fig.\ref{fig:lloyds_5n_df})  & 5 (Fig.\ref{fig:lloyds_5nf_df}) \\ \hline
LUT               & 8935                            & 10522                           & 12136                           & 14427                           & 5034                            \\
Registers         & 185                             & 296                             & 333                             & 382                             & 553                             \\
Memory bits       & 648                             & 648                             & 648                             & 648                             & 540                             \\
RAMB36E1          & 10                              & 33                              & 37                              & 104                             & 41                              \\
RAMB18E1          & 1                               & 2                               & 2                               & 3                               & 3                               \\
DSPs              & 108                             & 108                             & 108                             & 108                             & 12                              \\
FMAX (MHz)        & 25                              & 37                              & 37                              & 42                              & 75                              \\
Latency (cycles)  & 1                               & 3                               & 3                               & 4                               & 7                               \\
Latency (ns)      & 40                              & 81                              & 81                              & 95                              & 93                              \\ %
\end{tabular}
\label{table:lloyds_resources}
\vspace{-2em}

\end{table*}
In the graphs, the input is provided in chunks of $6$ to limit the number of inputs bits on the FPGA.
Figure~\ref{fig:lloyds_1n_df} shows one node containing the entire algorithm, it consumes $18$ coordinates and $3$ initial cluster center coordinates in one clock cycle.
The \gls{sdfap} graphs in this paper contain source and sink nodes that provide or consume data but those are not synthesized.
The next clock cycle it delivers the $3$ updated cluster center points.
To perform one iteration of the Lloyds algorithm in $1$ clock cycle for $18$ input points and $3$ cluster points requires the resources shown in the second column of Table~\ref{table:lloyds_resources}.
One input edge has the same production and consumption pattern, hence there is no need to initialize a complete FIFO and controller, only some registers.
The input and output values are vectors of tuples containing coordinates as $18$-bit values, therefore the amount of memory bits required is $18 \times (18 + 18) = 648$.

\begin{figure}
  \vspace{-2em}
  \centering
  \subfloat[\centering\scriptsize{3 nodes}]{
    \centering
    \includegraphics[height=14em]{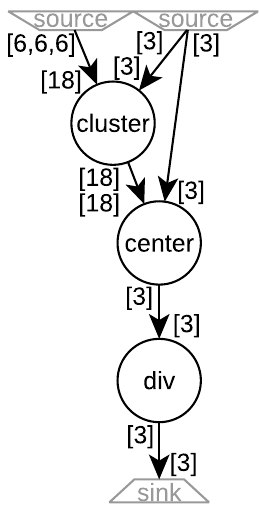}
    \label{fig:lloyds_3n_df}
  }
  \subfloat[\centering\scriptsize{center node extracted}]{
    \centering
    \includegraphics[height=14em]{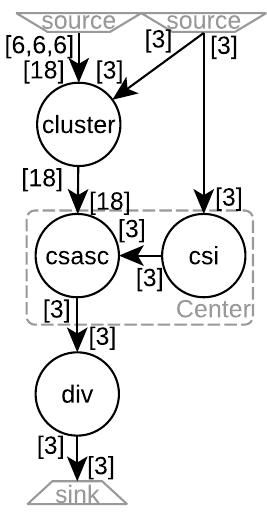}
    \label{fig:lloyds_4n_df}
  }
  \subfloat[\centering\scriptsize{cluster node extracted}]{
    \centering
    \includegraphics[height=14em]{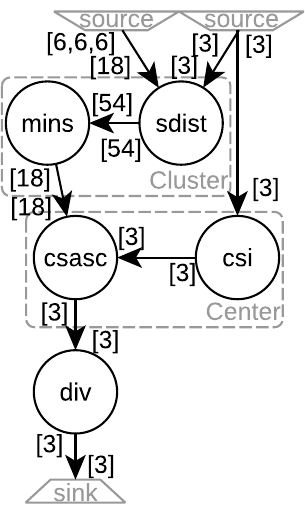}
    \label{fig:lloyds_5n_df}
  }
  \subfloat[\centering\scriptsize{longer aps}]{
    \centering
    \includegraphics[height=14em]{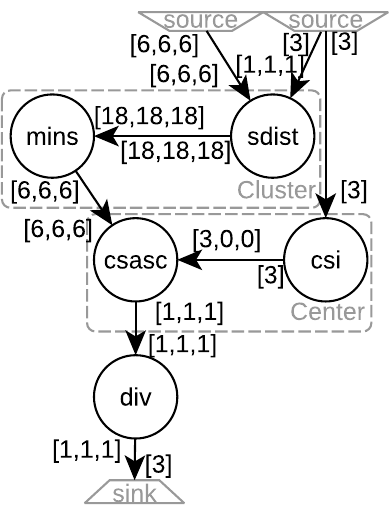}
    \label{fig:lloyds_5nf_df}
  }
  \caption{\gls{sdfap} graphs of Lloyds algorithm}
  \label{fig:lloyds_df}
  \vspace{-2em}
\end{figure}

Suppose we want to reduce the resource usage, then the straightforward solution would be to just adjust the patterns of the \texttt{lloyds} node.
However, there are foldable functions inside the \texttt{lloyds} node.
The \gls{hls} tool is currently unable to determine what the internal state must be if these foldable functions are somewhere inside the \gls{ast}.
Therefore, we need to decompose the nodes first, so that our tool can recognize the foldable functions, and hence determine what the internal state of that specific foldable function should be.
This limitation should be solved in the future by introducing a hierarchy on both clocked and unclocked input specifications.
For now, to reduce the resource usage, we need to decompose the nodes into smaller ones, until we have singled out all the foldable nodes.
Figure~\ref{fig:lloyds_3n_df} shows the Lloyds algorithm in \gls{sdfap} but now split into $3$ nodes, which results in a pipelined version of the algorithm.
The input from the source nodes provides the same data to the \texttt{cluster} node as well as to the \texttt{center} node.
The amount of LUTs increased due to the extra registers required for the additional edges.
Only registers are required because the production and consumption patterns for these edges are the same.
The splitting of nodes also results in a $48\%$ higher maximum clock frequency.
The throughput is still $1$ point every clock cycle, but the latency is increased to $3$ clock cycles.

In Figure~\ref{fig:lloyds_4n_df} the node that calculates the new center points is split into two nodes; \texttt{csasc} and \texttt{csi}.
Due to the additional edges, additional registers are required and hence the increase of LUTs (See Table~\ref{table:lloyds_resources}).
The \texttt{csasc} node is a foldable node that contains the \textit{foldl} function (See line 5).
The amount of \gls{dsp} blocks remains $108$.

Figure~\ref{fig:lloyds_5n_df} shows the node \texttt{cluster} also decomposed into $2$ separate nodes.
Again, introducing register consumption, increasing latency, but the throughput stays the same (Column 5 of Table~\ref{table:lloyds_resources}).
In this case, the \texttt{mins} node contains the foldable function \textit{foldl} (See Listing~\ref{lst:lloyds_def} line 3).
Now that all the foldable functions are in separate nodes we can start the time-area trade-off by changing the patterns.
\vspace{-.5em}
\begin{lstlisting}[label={lst:lloyds_def}, caption={node definitions of Figure~\ref{fig:lloyds_5nf_df}}, deletekeywords={foldl, div}]
tSdist=$(tool 'sdist [([6,6,6],[6,6,6]),([3],[1,1,1])]
                       [18,18,18])
tMins =$(tool 'mins  [([18,18,18],[18,18,18])]
                       [6,6,6])
tCsi  =$(tool 'csi   [([3],[3])]
                       [3])
tCsasc=$(tool 'foldl [([3],[3,0,0]),([6,6,6],[6,6,6])]
                       [1,1,1])
tDiv  =$(tool 'div   [([1,1,1], [1,1,1])]
                       [1,1,1])
\end{lstlisting}

\begin{lstlisting}[label={lst:lloyds_comp}, caption={nodes composed of Figure~\ref{fig:lloyds_5nf_df}}, deletekeywords={foldl, div}]
tCluster ps cs = mns where
  dys = tSdist ps cs
  mns = tMins dys

tCenter pscs cs = csast where
  cst = tCsi g cs
  csast = tCsasc f cst pscs

lloyds ps cs = cs' where
  pscs  = tCluster ps cs
  csast = tCenter pscs cs
  cs'   = tDiv csast
\end{lstlisting}

In Figure~\ref{fig:lloyds_5nf_df} the consumption and production patterns are changed so that the computations per node are divided over $3$ clock cycles.
As expected, the number of \gls{dsp} blocks required is lowered to $12$, also the logic utilization is roughly a third.
The foldable nodes now require an internal state, this state is stored in registers, hence the increase in the number of registers.
Due to the reuse of hardware over time the total amount of LUTs is also one-third of the LUTs used in the previous version.
The amount of blockRAM required is slightly lower compared to the previous version due to the changed consumption pattern on the \textit{sdist} node.
The code for the entire \gls{sdfap} graph is shown in Listing~\ref{lst:lloyds_def}.
Lines $1-10$ show the timed node definitions using the Template Haskell tool, for example, the \textit{mins} function is purely combinational that calculates the minimum value over a set of vectors.

Listing~\ref{lst:lloyds_comp} shows the composition of the nodes.
The \textit{tMins} is the generated clocked version of \textit{mins} with the desired input and output patterns.
Lines $1-3$ are the functional description of the \gls{sdfap} actor \texttt{cluster}, which is decomposed into $2$ nodes.
Lines $5-7$ are the description of the actor \texttt{center} and lines $9-12$ describe the composition of the \gls{sdfap} graph.
This section demonstrates the change in resource usage when nodes are decomposed and which introduces new edges.
It also highlights that support for hierarchy in the design specification can be desirable.

\section{Case studies}
\label{sec:case_study}

For evaluation and comparison, we implemented several algorithms using our proposed \gls{hls} method as well as the commercially available Vitis \gls{hls}, provided by Xilinx, currently the largest FPGA vendor.
For the dot-product case study, we also have a comparison with the \gls{hls}-tool provided by Intel.
We kept the input description for our tool and the Vitis tool as similar as possible to enable a fair comparison in terms of transparency, consistency, and performance of the resulting architecture.
As a consequence, we did not perform code transformations by hand, such as combining nested loops into a single loop.
All repetition in the C++ specification is expressed using for-loops and in the functional specification using \gls{hof}s.
One major difference between both approaches is that Vitis generates without pragmas a hardware architecture in which all computations are performed sequentially, whereas our \gls{hls}-tool generates without patterns a fully parallel combinational architecture.
For Vitis, we used pragmas, like unrolling and partitioning of data, to steer the tool.
For our \gls{hls}-tool, we used the access patterns.
For the synthesis of the generated Verilog code, we used Vivado v2020.2 and a Virtex 7 as target FPGA.

The algorithms we used for the case study are the dot-product, \gls{com} computations on images, and a 2D \gls{dct}.
The dot-product serves as a simple starting point and allows easy comparison of different versions.
The \gls{com} case study is slightly more complex but has dependencies that are straightforward to derive and has a lot of potential parallelism.
The 2D \gls{dct} has nodes that have fixed patterns because it has predefined IP blocks and hence those are modelled with fixed patterns.

The \gls{sdfap} graphs in these studies contain \texttt{source} and \texttt{sink} nodes that provide or consume data but those are not synthesized.
For a single node with patterns of length $1$, the overhead is in the range of $11$ LUTs with $11$ registers.
We also measured the time it took to generate Verilog from the input description with the \gls{hls}-tools.
Our \gls{hls}-tool took $48$ seconds to generate Verilog for all the designs, Vitis took $66$ minutes and $17$ seconds.

\subsection{Dot product case study}
\begin{wrapfigure}{r}{0.31\textwidth}
  \vspace{-2.5em}
  \centering
  \includegraphics[width=0.3\textwidth]{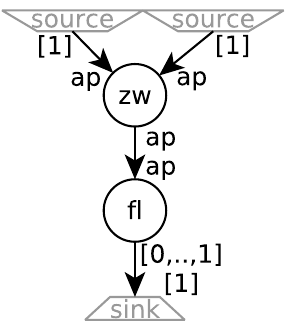}
  \caption{{\gls{sdfap} graph for dot product}}
  \label{fig:zwfl_ap_df}
  \vspace{-2em}
\end{wrapfigure}

The \gls{sdfap} graph of the dot-product is shown in Figure~\ref{fig:zwfl_ap_df}.
As mentioned in Section~\ref{sec:conformance_relation}, for every edge our tool synthesizes a FIFO, except the edges to the \texttt{sink} nodes.
A FIFO in our tool can either be a collection of registers or blockRAM.
\begin{table*}
  \vspace{-2em}
\centering
\caption{Resources usage of our HLS-tool in comparison with Intel HLS synthesized in Quartus 18.1}
\begin{tabular}{l||c|c|c|c||c|c|c|c|}
Dotproduct        & \multicolumn{4}{c||}{Our HLS-tool}                  & \multicolumn{4}{c|}{Intel HLS}               \\ \hline
ap                & {[}1,1,...,1{]} & {[}5,5,5,5{]} & {[}10,10{]} & {[}20{]} & no unroll & unroll 5 & unroll 10 & unroll 20 \\
ALMs              & 94              & 198           & 289         & 459      & 1701      & 4128     & 7731      & -         \\
Registers         & 67              & 73            & 77          & 84       & 2759      & 5901     & 10763     & -         \\
Memory bits       & 720             & 720           & 720         & 720      & 0         & 10240    & 20480     & -         \\
DSPs              & 1               & 3             & 6           & 10       & 1         & 3        & 5         & -         \\
FMAX (MHz)        & 199             & 174           & 170         & 170      & 206       & 197      & 168       & -         \\
Latency (cycles)  & 21              & 5             & 3           & 2        & 40        & 40       & 40        & -         \\
Latency (ns)      & 106             & 29            & 18          & 12       & 194       & 203      & 238       & -
\end{tabular}
\label{table:dotp_resources_quartus}
\vspace{-2em}

\end{table*}

For the Dot product case study we also made a comparison between our tool and the Intel \gls{hls}.
The nodes are described in the Intel C++ input as two for-loops and for our \gls{hls}-tool as two \gls{hof}s.
The synthesized results from Quartus 18.1 are shown in Table~\ref{table:dotp_resources_quartus}.
From the analysis of the results, we can conclude that in general the architecture produced by Intel \gls{hls} consumes significantly more resources.
This was because it generates a processor architecture using an Avalon bus system for all the communication.
Intel was unable to generate an architecture that showed correct behavior for the \textit{unroll 20} version.
For both tools, an increase in \gls{alm}s can be seen as we increase the parallelism.
In our \gls{hls} the register and blockRAM usage stay roughly the same.
The amount of blockRAM can be calculated for the two input edges that need to both store $20 \times 18$-bit values $=720$.
The edge between the \textit{zw} and \textit{fl} have the same production and consumption pattern and registers are introduced instead of a FIFO with a controller.
For Intel \gls{hls} both registers and blockRAM usage increases with a larger unroll.
The number of \gls{dsp}s increases also as more parallelism is introduced and Quartus can synthesize 1 \gls{dsp}s for 2 multiplications.
As parallelism increases in our \gls{hls}, we see a slightly lower FMAX due to the increased combinational path for the adders in the \textit{fl} node.
With Intel \gls{hls} we see the opposite and the FMAX increases as more parallelism is introduced.
The latency in nanoseconds that our tool can achieve is lower for all cases compared to Intel.

On average our tool uses $22$ times fewer resources and has a $7.5$ times lower latency both in clock cycles and in nanoseconds.
From a resource consumption, both in \gls{alm}s and blockRAM and latency perspective, our \gls{hls}-tool is more consistent compared to Intel \gls{hls}.
This case study also demonstrates that pragmas steering parallelism does not always predictably affect latency.
Since the Intel \gls{hls} tends to introduce a large bus and overhead, we decided to use Vitis \gls{hls} and Vivado for all the case studies.
For a comparison with Vitis, we used the same input specification for our \gls{hls} as we used for the comparison with Intel.
Vitis generates multiple input buffers in a sequence before streaming the values to the multipliers and adders.
It also creates pipelined adders to sum the results of the multiplications.
Both of these decisions result in an increased latency in nanoseconds, but allow for a higher clock frequency.
Table~\ref{table:dotp_resources_vivado} shows the comparison of resource conpsumption between our \gls{hls} and Vitis.
\begin{table*}
  \vspace{-2.5em}
\centering
\caption{Resources usage of our HLS-tool in comparison with Vitis HLS synthesized in Vivado}
\begin{tabular}{l||c|c|c|c||c|c|c|c|}
Dotproduct        & \multicolumn{4}{c||}{ Our HLS-tool}                  & \multicolumn{4}{c|}{Vitis HLS}               \\ \hline
ap                & {[}1,1,...,1{]} & {[}5,5,5,5{]} & {[}10,10{]} & {[}20{]} & no unroll & unroll 5 & unroll 10 & unroll 20 \\
LUTs              & 126             & 329           & 582         & 595      & 277       & 381      & 471       & 707       \\
Registers         & 66              & 261           & 449         & 1530     & 425       & 529      & 643       & 640       \\
RAMB18E1          & 1               & 0             & 0           & 0        & 0         & 0        & 0         & 0         \\
DSPs              & 1               & 5             & 10          & 20       & 1         & 5        & 10        & 20        \\
FMAX (MHz)        & 150             & 162           & 146         & 131      & 283       & 189      & 148       & 107       \\
Latency (cycles)  & 21              & 5             & 3           & 2        & 29        & 42       & 57        & 49        \\
Latency (ns)      & 140             & 31            & 21          & 15       & 103       & 222      & 386       & 459
\end{tabular}
\label{table:dotp_resources_vivado}
\vspace{-2em}
\end{table*}
The amount of LUTs and registers increases with introducing more parallelism.
When the FIFO depth is below a certain threshold Vivado introduces registers instead of blockRAM for storage.
The register usage for our \gls{hls} increases with more parallelism.
\gls{dsp} usage scales according to the parallelism introduced using patterns or pragmas.
The FMAX decreases as more parallelism is introduced but a steeper decrease is shown for Vitis.
Latency in cycles shows unpredictable behavior for Vitis.
Somehow the tool is unable to find a shorter schedule for the extra parallelism that is introduced.
This results in much longer latency in nanoseconds compared to our \gls{hls}-tool.
The architectures produced by Vitis achieve a higher FMAX, except the \textit{unroll 20} variant, but combined with the latency in cycles the latency in nanoseconds is significantly longer compared to our \gls{hls}-tool, except the \textit{no unroll} variant.

On average Vitis requires $30\%$ more LUTs and has a $14$ times higher latency in nanoseconds compared to the architectures produced by our \gls{hls}-tool.
From a resource consumption and latency perspective, our \gls{hls}-tool is more consistent compared to Vitis.
This case study also demonstrates that pragmas steering parallelism does not always predictably affect latency.

\subsection{Center of Mass case study}
\begin{wrapfigure}{}{0.22\textwidth}
  \vspace{-1em}
  \centering
  \includegraphics[width=.20\textwidth]{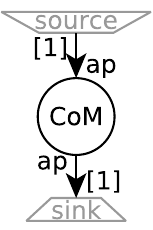}
  \caption{\gls{sdfap} graph for CoM}
  \label{fig:com_1n}
  \vspace{-2em}
\end{wrapfigure}

For this case study, we use a center of mass computation on gray-scale images to demonstrate the effect of parallelization through patterns or pragmas.
An image is chopped into blocks of $8 \times 8$ pixels and the center of mass of those blocks is computed.
Vitis synthesizes a large input buffer where the pixels are streamed into.
From this buffer the data is fed through a pipelined version of the algorithm consisting of adders and multipliers.
The computation for the center of mass of an $8 \times 8$ image does not require many \gls{dsp}s.
As shown in the first column of Table~\ref{table:com_resources}, calculations for an $8 \times 8$ image do not require many resources and hence we can parallelize to decrease latency in nanoseconds.
In this column, we set the ap to $[1]$ to reflect a single \gls{com} calculation for an $8 \times 8$ image.
For our \gls{hls} the number of logic cells and \gls{dsp}s is consistent with the parallelism specified by the pattern.
For a single computation, $20$ \gls{dsp}s are required, when we use the access pattern $ap = [16,16,...,16]$, we need $16 \times 20 = 320$ \gls{dsp}.
For the access pattern $ap = [64,64,64,64]$ we need $64 \times 20 = 1280$ \gls{dsp}s.
For the access patterns $[16,16,...,16]$ and $[64,64,64,64]$ Vivado introduces blockRAM to store data.
The FMAX stays roughly constant because the longest combinational path is not increased, only extra parallelism is introduced.
Since there is a single node in Figure~\ref{fig:com_1n}, the latency in clock cycles is the length of the pattern.

With Vitis, we also see an increase in LUTs when more parallelism is introduced using pragmas.
From the \textit{unroll 16} variant to the \textit{unroll 64} variant no extra blockRAMs or \gls{dsp}s are introduced, but many more registers.
The latency in cycles also stays at $700$, meaning that Vitis is unable to find a better schedule compared to the \textit{unroll 16} variant.

The patterns show a consistent and predictable behavior in terms of resource consumption and latency.
The Vitis tool is not able to generate a schedule that efficiently utilizes the extra parallelism introduced using pragmas.
On average the Vitis architecture consumed $40\%$ fewer LUTs and $15$ times more registers.
Vitis can achieve a higher clock frequency but a large number of clock cycles results in an on average $18$ times higher latency in nanoseconds compared to our \gls{hls}.
\begin{table*}
  \vspace{-2.5em}
\centering
\caption{Resources usage for different versions of the \gls{com} algorithm}
\begin{tabular}{l||c|c|c||c|c|c|}
CoM               & \multicolumn{3}{c||}{Our HLS}                        & \multicolumn{3}{c|}{Vitis HLS} \\ \hline
ap                & $[1]$     & $[16,16..,16]$ & $[64,64,64,64]$  & no unroll & unroll 16 & unroll 64 \\
LUTs              & 1577      & 31744          & 125017           & 2035      & 11513     & 29476     \\
Registers         & 681       & 355            & 1026             & 2438      & 8309      & 19783     \\
RAMB36E1          & 0         & 144            & 576              & 0         & 0         & 0         \\
RAMB18E1          & 0         & 0              & 0                & 66        & 132       & 132       \\
DSPs              & 20        & 320            & 1280             & 12        & 24        & 24        \\
FMAX (MHz)        & 33        & 28             & 27               & 157       & 115       & 110       \\
Latency (cycles)  & 256       & 16             & 4                & 825       & 700       & 700       \\
Latency (ns)      & 7724      & 565            & 151              & 5265      & 6096      & 6366
\end{tabular}
\label{table:com_resources}
\vspace{-2em}
\end{table*}

\subsection{DCT2D case study}
\begin{figure}
  \vspace{-1.5em}
  \centering
  \includegraphics[width=\textwidth]{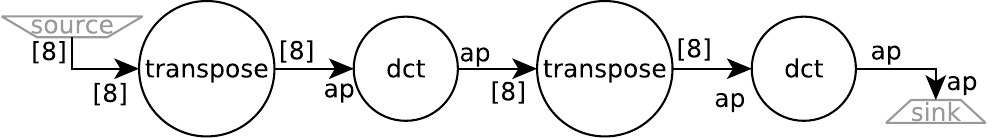}
  \caption{\gls{sdfap} graph for a 2D \gls{dct}}
  \label{fig:sdfap_dcd1d_ap}
  \vspace{-2em}
\end{figure}

The \gls{dct} is implemented using an $8 \times 8$ input matrix with 18-bit values.
The \gls{sdfap} graph is shown in Figure~\ref{fig:sdfap_dcd1d_ap} where the access pattern \textit{ap} is shown as a variable.
The numbers in the patterns represent the number of vectors containing $8$ 18-bit values, so $[4,4]$ means an input width of a $ 4 \times 8$ matrix.
The patterns for the \texttt{transpose} nodes remain fixed to demonstrate the case when the engineer is not able to change the behavior of certain nodes (which is the case for external IPs).
An overview of the hardware synthesis of different access patterns for the three different FIFO types is shown in Table~\ref{table:dct2d_resources}.
\begin{table*}
  \vspace{-1.5em}
  \vspace{-2em}
\centering
\caption{Resources usage for different versions of the DCT2D algorithm}
\begin{tabular}{l||c|c|c|c||c|c|c|c|}
DCT2D             & \multicolumn{4}{c||}{Our HLS-tool}       & \multicolumn{4}{c|}{Vitis HLS}                 \\ \hline
ap                & [1,1,...,1] & [2,2,2,2]  & [4,4] & [8]   & no unroll & unroll 2  & unroll 4  & unroll 8 \\
LUTs              & 2569        & 4224       & 7786  & 17442 & 3370      & 2622      & 3507      & 4558     \\
Registers         & 5308        & 5311       & 5311  & 5311  & 4577      & 5108      & 5091      & 6083     \\
RAMB18E1          & 0           & 0          & 0     & 0     & 5         & 19        & 21        & 8        \\
DSPs              & 56          & 112        & 224   & 448   & 56        & 4         & 16        & 28       \\
FMAX (MHz)        & 91          & 92         & 98    & 102   & 171       & 205       & 189       & 148      \\
Latency (cycles)  & 18          & 10         & 6     & 4     & 213       & 2322      & 2162      & 269      \\
Latency (ns)      & 199         & 109        & 61    & 39    & 1246      & 11327     & 11439     & 1818
\end{tabular}
\label{table:dct2d_resources}
\vspace{-2em}
\end{table*}

The number of logic cells used roughly doubles when we double the parallelism using patterns.
This amount of registers stays roughly the same through all the patterns since we do not introduce new edges.
The number of \gls{dsp}s is consistent with the specified access patterns, scaling the pattern with the factor 2 also doubles the \gls{dsp} consumption.
The longest combinational path is not changed by parallelization and hence the FMAX remains almost the same.
The latency in cycles is deduced from the \gls{sdfap} graph and the latency in nanoseconds shows a predictable decrease when more parallelism is introduced.

Vitis is somehow able to utilize more parallelism if no pragmas are given, the tool does not tell us how and why this is the case.
This parallelization effect is especially visible in the \gls{dsp} consumption and the latency in cycles.
However, when we signal the compiler that some of the loops can be unrolled, it introduces more LUTs for the \textit{unroll 4} and \textit{unroll 8} variant.
The latency in cycles shows very inconsistent behavior since the \textit{unroll 2} and \textit{unroll 4} variants have a latency that is 10 times higher than the \textit{no unroll} variant.

From the results of Table~\ref{table:dct2d_resources}, we conclude that the usage of logic cells for our \gls{hls}-tool on average is $30\%$ higher and that the \gls{dsp} usage scale predictable according to the specified input patterns.
The speed-up for our tool varies between $6.3$ and $188$ times and is $86$ on average.

\section{Conclusion}
\label{sec:conclusion}
We combined the \gls{sdfap} temporal analysis model with a functional input language to automatically generate both control and datapath in hardware.
Access patterns are used to specify resource usage and temporal behavior, providing the engineer with a transparent way of performing the time-area trade-off.
From the schedules of these \gls{sdfap} graphs follow the latency and throughput of the generated hardware.

Our \gls{hls}-tool uses the metaprogramming capabilities of Template Haskell to modify the \gls{ast} during the compilation process.
The existing Clash-compiler is then used to generate VHDL or Verilog.
Invalid patterns can be detected in the early stages of compilation and the process can be stopped and the engineer notified.
The amount of control hardware overhead depends on the chosen design granularity but is overall small.
For a single node with patterns of length $1$, the overhead is in the range of $40$ \gls{alm}s and $50$ registers.

Access patterns in our \gls{hls}-tool offer much more control over the resulting architecture compared to the pragmas of Vitis.
Doubling pattern length while halving the consumption, results in half the \gls{dsp} and logic cell consumption, but double the latency.
Using \gls{sdfap} opens up the possiblity of empolying dataflow analysis techniques.
Case studies show consistent resource consumption and temporal behavior for our \gls{hls}.

Resource consumption in the Dot product case study is $30\%$ lower compared to Vitis and the average speedup in latency in nanoseconds is $14$ times.
For the 2D DCT, our \gls{hls}-tool utilizes $30\%$ more LUTs but is $86$ times faster on average.
For the \gls{com} case study our tool consumed on average $40\%$ more logic cells but can achieve a speedup of $18$ times.
For both the 2D DCT and the \gls{com} case study Vitis is unable to utilize the extra parallelism introduced with pragmas and resulting in an inefficient schedule that leads to high latency in nanoseconds.

\vspace{-1em}

\bibliographystyle{splncs04}
\bibliography{bibliography}

\end{document}